# Ab Initio Study of Phase Stability in Doped $TiO_2$

**Dorian A.H. Hanaor[*], Mohammed H.N. Assadi, Sean Li, Aibing Yu and Charles C. Sorrell**

School of Materials Science and Engineering, University of New South Wales, Sydney, NSW 2052, Australia

**Abstract:**

Ab-initio density functional theory (DFT) calculations of the relative stability of anatase and rutile polymorphs of $TiO_2$ were carried using all-electron atomic orbitals methods with local density approximation (LDA). The rutile phase exhibited a moderate margin of stability of ~ 3 meV relative to the anatase phase in pristine material. From computational analysis of the formation energies of Si, Al, Fe and F dopants of various charge states across different Fermi level energies in anatase and in rutile, it was found that the cationic dopants are most stable in Ti substitutional lattice positions while formation energy is minimised for $F^-$ doping in interstitial positions. All dopants were found to considerably stabilise anatase relative to the rutile phase, suggesting the anatase to rutile phase transformation is inhibited in such systems with the dopants ranked F>Si>Fe>Al in order of anatase stabilisation strength. Al and Fe dopants were found to act as shallow acceptors with charge compensation achieved through the formation of mobile carriers rather than the formation of anion vacancies.

Keywords: Solar Energy, Titanium Dioxide, Photocatalysis, Density Functional Theory

## 1. Introduction

Over the last 4 decades, titanium dioxide has been attracting significant research interest in the field of renewable energy. Owing to the particular energetic levels of its conduction and valence bands, titanium dioxide is able to function as a semiconductor photocatalyst and facilitate reactions involved in applications including hydrogen production through water splitting (solar hydrogen) [1-4], sunlight-driven water purification [5-8] , self cleaning coatings [9-12] and self sterilizing coatings [13-15]. These applications are of great interest in facilitating environmental remediation and in enabling the conversion of solar energy to a storable medium. Of further interest in the field of renewable energy generation, $TiO_2$ is also of great promise in photovoltaic applications and is widely used for the fabrication of photo-anodes in Dye Sensitized Solar Cells (DSSCs) [16-19].

Titanium dioxide, as with other semiconductor photocatalysts, facilitates reactions through the photo-generation of electron-hole pairs (excitons) by irradiation exceeding the material's band gap. These electron-hole pairs tend to undergo rapid recombination, a phenomenon which can be observed through photoluminescence, [20, 21] or alternatively, can react with surface adsorbed species in order to directly or indirectly facilitate desired reactions [22-24].

Titanium dioxide is commonly used in its two main phases, the equilibrium rutile phase and the metastable anatase phase, which transforms to rutile through thermal treatment. Although rutile exhibits a narrower band-gap than anatase, ~3.0eV compared with ~3.2eV, [25-27] anatase is generally considered to exhibit superior photocatalytic performance, owing to higher levels of surface area and thus higher activity [28].

Owing to a more flexible assembly of 4-edge-sharing $TiO_6$ octahedra, anatase, rather than rutile, is frequently the first crystalline phase formed in many synthesis routes. [29, 30] Mixed phase compositions of anatase and rutile are widely accepted to exhibit improved exciton separation, through a trapping of conduction band electrons in the rutile phase and valence band holes in the anatase phase, and consequently superior photocatalytic activity as well as superior performance in photovoltaic applications [31-36]. A third metastable phase, brookite, has also been reported to exhibit photocatalytic activity [37, 38], and indeed mixtures of anatase rutile and brookite have been reported to exhibit good levels of activity[38, 39]. However, this phase is of lesser interest in energy applications owing to the complexity in its synthesis.

The application of titanium dioxide in photocatalysis has been drawing increasing scientific interest over recent years and many studies have focused on enhancing the performance of $TiO_2$ photocatalysts. In essence, the enhancement of the activity of $TiO_2$ photocatalysts is attempted through the enhancement of exciton generation, and/or the reduction of exciton recombination.

The addition of anionic and cationic dopants is a common method of controlling the properties of titanium dioxide and enhancing the material's performance as a photocatalyst. Dopant elements can be added to $TiO_2$ to achieve various outcomes

- Formation of new valence states in $TiO_2$ [40-43]
- Creation of charge carrier trapping sites [44-47]
- Band gap reduction [48-50]
- Control of phase transformation behaviour [51-54]
- Surface area enhancement [53, 55]

In particular, dopants, or unintentional impurities, have a pronounced effect on the anatase to rutile phase transformation. Certain dopants are reported to promote the anatase to rutile phase transformation, allowing this reconstructive process to occur at lower temperatures and take place more rapidly. Conversely, dopants are often reported to inhibit the anatase to rutile phase transformation, imparting greater stability to the anatase phase. The effects of dopants on the anatase to rutile phase transformation have been reviewed comprehensively elsewhere. [29] Briefly, the promotion of the anatase to rutile phase transformation is generally reported to occur through an easing of the atomic rearrangement involved in the transformation, often as the result of an increase in the density of anion vacancies. This may occur when cationic dopants of low valence substitute for Ti in the anatase lattice [29, 56, 57]. Conversely, the inhibition of the anatase to rutile phase transformation is accepted to occur through a restriction of the atomic rearrangement involved in the phase transformation. This may occur as the result of the substitution of $Ti^{4+}$ with cations of equal or higher valence, or through the presence of dopant elements located in interstices or at grain boundaries. Although the effects of dopants on the anatase to rutile phase transformation can ostensibly be predicted from consideration of ionic radii and valence, some controversy exists as numerous dopants are reported both as inhibitors and as promoters of the phase transformation and the mechanisms through which doping with such elements takes place and through which the phase transformation kinetics are altered remain unclear. The ambiguity regarding the effects of certain dopants on the anatase to rutile phase transformation may stem from the sensitivity of the material to experimental conditions.

In the present work we aim to shed light on the location of dopant elements by examining the role of sole-dopants from an energetics perspective through computer simulation of the formation energies of dopants in substitutional and interstitial positions. The effects of dopant elements on the stability of the anatase phase is predicted by the analysis of total free energy of doped $TiO_2$ in anatase and rutile phases. An understanding of the fundamental effects of dopants on phase stability and transformations in $TiO_2$ is of great importance for the use of this material in energy applications.

## 2. Methodology

First-principles density functional theory (DFT) enables the calculation and prediction of material properties directly from quantum mechanical considerations, without the aid of phenomenological parameters. DFT methods evaluate the electronic structure by constructing a potential acting on a system's electrons. The DFT potential is a sum of external potentials ($V_{ext}$) which is exclusively determined by the geometry and the chemical composition of the system, and an effective potential ($V_{eff}$) that represents the inter-electronic interactions. As a result, a DFT problem for a system with $n$ electrons is a set of $n$ one-electron Schrödinger-like equations which are known as Kohn-Sham equations [58]

$$H\Psi_n = \left(-\frac{\hbar^2}{2m}\nabla^2 + V_{ext} + V_{eff}\right)\Psi_n = \varepsilon_n \Psi_n \qquad (1)$$

Here, $\Psi_n$ are the n one-electron wavefunctions, $V_{ext}$ is the external potential of the nuclei, $V_{eff}$ the effective potential, $-(\hbar^2\nabla^2/2m)$ is the kinetic energy operator and $\varepsilon_n$ is the DFT eigenvalue. $V_{eff}$ contains terms that govern the electronic exchange and correlation ($V_{xc}$) which are usually approximated by various formalisms, which have been reviewed elsewhere [59] Since the exchange energy is better known and characterized for uniform electrons, from a computational point of view, it is practical to approximate the exchange-correlation energy terms of any system by the one of the uniform electronic gas. This procedure is called Local Density Approximation (LDA) and offers a very popular and cost-effective method for conducting DFT calculations [60].

### 2.1. Computational Settings

*Ab initio* calculations were performed with the use of DMol$^3$, a density functional theory based software package developed by Accelrys [61, 62]. Here a spin unrestricted method was utilized where spin-up and spin down electrons were considered independently to allow accurate calculations of spin related phenomena. This is important for transition elements where the d orbitals are partially filled and spin distribution over the orbitals is asymmetric. The *Double-numeric plus polarization* (DNP) basis set was used to construct the molecular orbitals. The chosen basis set gives an accurate description of bonding and is generally more reliable than DMOL$^3$'s other available sets such as the minimal basis set which are often are inadequate for quantitative analyses.

Local density approximation (*LDA*) based on Perdew-Wang formalism was applied to approximate the exchange-correlation energy [63]. Here $V_{xc}$ was calculated from the DFT total energy ($E^{LDA}$) according to the standard LDA procedure shown in following equation [64]:

$$V_{xc}^{LDA}(r) = \frac{\delta E^{LDA}}{\delta \rho(r)} \qquad (2)$$

where the exchange-correlation potential ($V_{xc}$) was directly related to the electronic density $\rho(r)$ at point *r*.

Real-space global cutoff radii were set at 0.52 nm for all elements and thus beyond that distance the wave functions of all elements were assumed to be zero. This allows the mathematical convergence of the calculated wave-functions without significantly compromising the accuracy of the calculation. The value of 0.52 nm for real-space global cut-off radii has been demonstrated to be adequate for metal oxide DFT calculations [61].

Since periodic boundary conditions were imposed during the calculation, Bloch's theorem [65] was applicable to the calculations and provided a great reduction in the required computational resources. Since electronic states are only allowed at a set of *k*-points, determined by the periodic boundary conditions, the infinite number of electrons in the periodic solid are accounted for by an infinite number of *k*-points. Bloch's theorem changes the problem of calculating an infinite number of electronic wavefunctions to one of calculating a finite number of wavefunctions at an infinite number of *k*-points confined to the first Brillouin zone. Since the electronic wavefunctions at *k*-points that are very close together vary very smoothly, DFT expressions that contain a sum over *k*-points, such as eigenevalues, can be efficiently evaluated using a numerical scheme that performs summation over a small number of special points in the Brillouin zone. In the present work, Brillouin zone sampling was carried out by choosing a *k*-point set generated by the Monkhorst-Park scheme with grid spacing of ~0.2 nm$^{-1}$ between *k*-points for all studied configurations and the total energy $E^{total}$ in the system was determined by integrating the wave function over the *k*-space [66]. For verification purposes, convergence testing was performed, first by increasing *k*-point density and then by increasing the real space global cut-off radii; it was found that the total energy differs less by 10$^{-5}$ eV/atom. Thus the results are considered well converged.

For geometry optimization, both DFT total energy and its first derivatives, with respect to ionic coordinates, are minimized by the displacement of ionic positions. The derivatives of the total energy with respect to the ionic positions, shown in Eq. 3, are termed Hellman Feynman forces, and in equilibrium should equal zero [67].

$$\vec{F}_i = -\frac{\partial E^{total}}{\partial \vec{r}_i} \qquad (3)$$

Minimizing $E^{total}$ and $\vec{F}_i$ with respect to the ionic coordinates ($\vec{r}_i$) is a commonplace optimization problem, and in DMOL$^3$ a standard Broyden–Fletcher–Goldfarb–Shanno method, also known as the BFGS algorithm, [68] is utilized iteratively for this purpose. DFT methods based on this algorithm have been widely employed in similar studies and are described in greater detail elsewhere [69, 70].

The convergence thresholds for energy, Cartesian components of internal forces acting on the ions, and displacement values were set to be 10$^{-5}$ eV/atom, 0.5 eV/nm, and 10$^{-6}$ nm, respectively. To avoid artificial hydrostatic pressure during the simulation, supercell lattice constants were fixed to the theoretical values and only internal coordinates were allowed to relax to a minimum energy configuration.

For doped systems, the formation energy ($E^f$) of each dopant in both substitutional and interstitial sites is calculated as function of Fermi level energy ($E_{Fermi}$). Then the formation enthalpies ($\Delta^F$) of the impurity doped, *n*-type rutile and anatase TiO$_2$ are compared to probe the effect of the dopants in anatase phase stabilisation.

Calculations were performed for pristine TiO$_2$, and TiO$_2$ doped with Si, Al, Fe, and F. This dopant range was selected as it represent dopants which are commonly found (although often unappreciated) as impurities in TiO$_2$.

- Si and Al are common cationic impurities that may derive from the glass and single crystal substrates commonly used in thin film fabrication
- Fe is a common cationic impurity, frequently used as a dopant in many materials and the interpretation of its effects on the anatase to rutile phase transformation is controversial.
- F$^-$ derives from fluorine doped tin oxide conducting substrates and it is a strongly electronegative anionic impurity/dopant.

## 3. Results

### 3.1. Pristine TiO$_2$

The optimized crystal structures, including lattice parameters, and the formation enthalpy ($\Delta^F$) of pristine rutile and anatase TiO$_2$ were determined by allowing all lattice constants and internal coordinates to relax as presented in **Table 1**. $\Delta^F$ was calculated according to the following formula [71]:

$$\Delta^F(TiO_2) = E^{total}(TiO_2) - E^{total}[Ti(metal)] - E^{total}[O_2(g)] \qquad (1)$$

where $E^{total}$ is the density functional theory (DFT) total energy of TiO$_2$, metallic Ti and O$_2$.

The calculated lattice parameters for a-TiO2 are a = 0.375 nm, c = 0.961 nm and u = 0.206 (u is the fractional *z* component of the oxygen position in TiO$_2$) . For r-TiO2, the calculated lattice parameters were found to be a =0.457 nm and c = 0.294 nm. All of these parameters are in good agreement with prior observations differing from the experimental values by less than 1% [72]. The difference in lattice parameters is consistent with the well- known bond softening effect of LDA functions [59]. The $\Delta_F$ of a-TiO$_2$ and r-TiO$_2$ were calculated to be -9.860 eV/f.u. and -9.863 eV/f.u. respectively which are in reasonable agreement with previous DFT calculations based on plane-wave pseudopotential methods.[73] In this work, $\Delta E$ is defined as $\Delta^F$(r-TiO$_2$) – $\Delta^F$(a-TiO$_2$) and is an indicator of the phase stability in TiO$_2$. Positive $\Delta E$ implies that a-TiO$_2$ is more energetically favourable over r-TiO$_2$. The presented DFT calculations indicate that, at standard conditions of temperature and pressure, in pristine TiO$_2$, r-TiO$_2$ is slightly more stable than a-TiO$_2$, as $\Delta E$ equals only to -3 meV. Consequently, such a small stability margin leaves the possibility of a-TiO$_2$ stabilization, depending on conditions of crystal growth and the presence of impurities. It should be noted that since the r-TiO$_2$ is more stable than a-TiO$_2$, the phase transformation from a-TiO$_2$ to r-TiO$_2$ is irreversible, as is widely reported in the literature [74, 75].

**Table 1. :Lattice parameters and formation enthalpy of anatase and rutile. Calculated values and experimental values reported in the literature [72]. (f.u.=formula unit)**

| Property | Anatase TiO2 | | Rutile TiO2 | |
|---|---|---|---|---|
| | Calculated | Experimental | Calculated | Experimental |
| a (nm) | 0.375 | 0.3784 | 0.457 | 0.45936 |
| c (nm) | 0.961 | 0.952 | 0.294 | 0.296 |
| u | 0.206 | 0.208 | - | - |
| $\Delta^F$ (eV/f.u.) | -9.860 | - | -9.863 | - |
| $\Delta^F$ (KJ/mole) | -951.345 | - | -951.634 | - |

### 3.2. Doped TiO$_2$

To explore the strategies to stabilize a-TiO$_2$ over r-TiO$_2$ by doping, the formation energy of various charge states of the Si, Al, Fe and F dopants in both a-TiO$_2$ and r-TiO$_2$ systems were calculated. Both substitutional and interstitial sites of both TiO$_2$ polymorphs were considered for impurity doping. **Figure 1**(a) and (b) schematically represent the lattice sites of the substitutional and interstitial dopants in a-TiO$_2$ and r-TiO$_2$, respectively. The formation energy of doping dopant was calculated according to the following formula: [76]

$$E^f = E^{total}(TiO_2:M) - E^{total}(TiO_2) - \mu_N + \mu_M + qE_{Fermi} \qquad (2).$$

Here, $E^{total}(TiO_2:M)$ is the DFT total energy of the doped system, $E^{total}(TiO_2)$ is the total energy of the undoped system, $\mu_N$ is the chemical potential of the substituted element, if any, $\mu_M$ is the chemical potential of the impurity and $q$ is the total charge dopant is has a positive (negative) values for donors (acceptors). $E_{Fermi}$ is the Fermi level energy with respect to the valence band maximum (VBM). There are number of factors that can affect the position of $E_{Fermi}$ in an oxide, including (a) carrier concentration, temperature, presence of co-dopants and unintentional charge compensators. However, TiO$_2$ is known to exhibits persistent n-type conductivity due to its metal excess nature which implies that is located near the conduction band maximum (CBM) edge. The chemical potentials of metallic elements, following common practice, [73, 77] were calculated from the DFT total energy of the corresponding metal oxide. For gaseous elements (O and F) the chemical potential was set to equal the molecule's total energy per element, thus reflecting the availability of single oxygen atoms (or fluorine) for the formation of the oxide [76].

Using the obtained formation energies ($E^f$), one can calculate the transition level, $\varepsilon(q/q')$, of each dopant which is defined as the Fermi level ($E_{Fermi}$) position where the charge states of a dopant $q$ and $q'$ have equal formation energy, and is calculated for a dopant as follows:

$$\varepsilon(q/q') = \frac{E^f(TiO_2:M,q) - E^f(TiO_2:M,q')}{q-q'} \qquad (3).$$

The experimental significance of the $\varepsilon(q/q')$ value is that for $E_{Fermi}$ positions below $\varepsilon(q/q')$ charge state $q$ is stable, while for $E_{Fermi}$ positions above $\varepsilon(q/q')$ charge state $q'$ is relatively stable. In other words, $\varepsilon(q/q')$ can determine the electrical activity of a dopant. If a dopant's $\varepsilon(q/q')$ is positioned within few meV of the band edges (valence-band maximum (VBM) for an acceptor, conduction-band minimum (CBM) for a donor), such a dopant is likely to be thermally ionized at room temperature and the dopant is called a shallow donor/acceptor. If $\varepsilon(q/q')$ is located far from the band edges, it is unlikely to be ionized at room temperature, and thus it constitutes a deep level dopant.

In calculations for all doped systems, a supercell of $3a \times 2a \times 2c$ relative to the primitive cell, containing 72 ions, was considered for both $TiO_2$ polymorphs. By introducing one dopant atom per supercell, one obtains a cation doping concentration of ~ 4.17 at% for dopants substituting a Ti ion and an anion doping concentration of ~ 2.08 at% for elements substituting an O ion. To examine the suitability of this supercell for accurate $E^f$ calculations, test calculations were performed on the r-$TiO_2$:$Si_{Ti}$ (rutile doped substitutionally with Si) system as a selected test system. The $E^f$ values of neutral $Si_{Ti}$ in as $3a \times 2a \times 2c$ and $3a \times 3a \times 2c$ and $3a \times 3a \times 3c$ were calculated to be 0.369 eV, 0.371 eV and 0.372 eV respectively, indicating only a very minor change in formation energy in larger supercells. Thus it can be concluded that a $3a \times 2a \times 2c$ supercell size is adequate for the purpose of calculating $E^f$. Calculated dopant formation energies in anatase and rutile and $\varDelta E$ values are shown in **Table 2**.

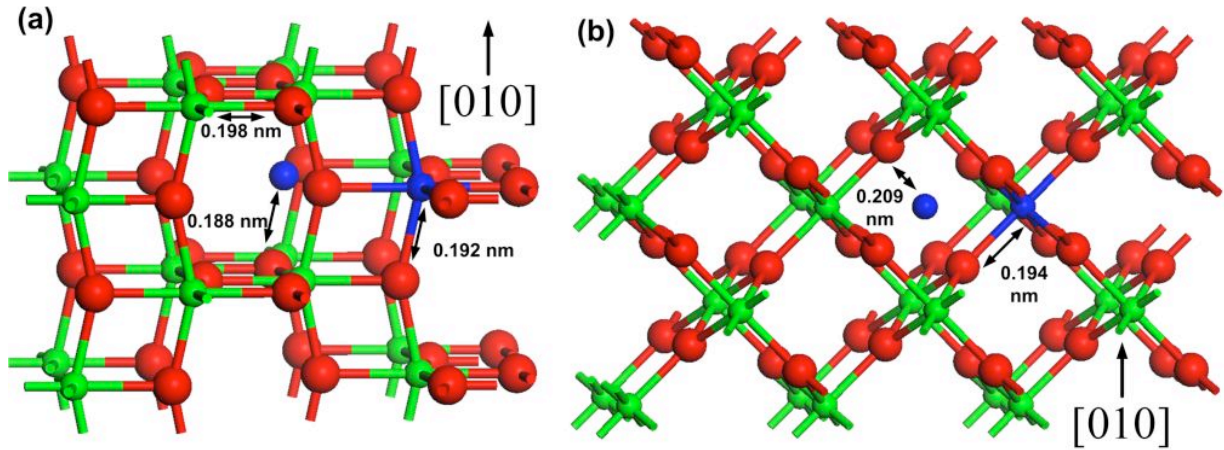

**Figure 1. Schematic representation of interstitial and substitutional dopant positions in (a) anatase and (b) rutile TiO$_2$ with arrows showing cation-dopant to oxygen distances.**

*TiO2:Si*

Si has two common oxidation states, i.e. Si$^{2+}$ and Si$^{4+}$. When a Si$^{2+}$ substitutes a Ti$^{4+}$ ($Si_{Ti}^{2+}$) in TiO$_2$'s host lattice, it leaves two holes in the valance band, while a $Si_{Ti}^{4+}$ is isovalent to Ti$^{4+}$ and has no electrical activity. For interstitial Si ($Si_{Int}$), $Si_{Int}^{0}$ is a neutral dopant, while $Si_{Int}^{2+}$ and $Si_{Int}^{4+}$ are double and quadruple donors, respectively. The $E^f$ of Si in a-TiO$_2$ and r-TiO$_2$ are presented in **Figure 2**(a) and (b) respectively. For a-TiO$_2$, $Si_{Ti}$ is more stable than $Si_{Int}$ for the entire range of the band gap with ε(4+/2+) = 1.439 eV, indicating that $Si_{Ti}$ is a deep double acceptor. Here, when the Fermi level is located at the VBM ($E_{Fermi}$ = VBM), that is p-type TiO$_2$, the $E^f$ of both $Si_{Ti}$ and $Si_{Int}$ is ~ 0.400 eV. However as Fermi level moves to the CBM as the case of n-type TiO$_2$, $Si_{Ti}^{2+}$ become more stable than $Si_{Int}^{0}$ by a large margin of 11.623 eV. In r-TiO2, when the Fermi level is positioned near CBM, $Si_{Int}^{4+}$ is more stable than $Si_{Ti}^{4+}$ by 2.935 eV. However at $E_{Fermi}$ = 0.826 eV , the $E^f$ of $Si_{Ti}^{4+}$ and $Si_{Int}^{4+}$ become equal. When $E_{Fermi}$ moves towards the CBM (as in *n*-type r-TiO$_2$) the $Si_{Ti}^{2+}$ form becomes more stable than $Si_{Int}^{0}$ by 8.742 eV, indicating that $Si_{Ti}$ is more stable than $Si_{Int}$ for *n*-type r-TiO$_2$. The transition level ε(4+/2+) for $Si_{Ti}$ is 1.821 eV. For *n*-type TiO$_2$, *ΔE* ($\left[\Delta^F(r-TiO_2:Si)-\Delta^F(a-TiO_2:Si)\right]/f.u.$) is 32.43 meV which indicates the Si dopant acts as an inhibitor to anatase to rutile phase transformation.

*TiO$_2$:Al*

Al usually occurs in 3+ charge state. In TiO$_2$ host lattice environment substitutional Al$^{3+}$ ($Al_{Ti}^{3+}$) acts as single acceptor while interstitial Al$^{3+}$ ($Al_{Int}^{3+}$) acts as triple donor. The $E^f$ of $Al_{Ti}^{3+}$ in a-TiO$_2$ is presented in **Figure 2**(c). It is evident that $Al_{Ti}$ is more stable than $Al_{Int}$ for the entire range of the band gap in anatase TiO$^2$. Additionally $Al_{Ti}^{3+}$ is always more stable than $Al_{Ti}^{4+}$ which is neutral with respect to the host crystal environment, indicating that Al is a shallow acceptor in a-TiO$_2$. For r-TiO$_2$, as in **Figure 2**(d), when the Fermi level is located at CBM, $E^f$ of $Al_{Int}^{3+}$ is -1.593 eV, lower than the one of $Al_{Ti}^{4+}$ which is 0.519 eV. As the Fermi level moves to VBM, $Al_{Ti}^{3+}$ becomes more stable than $Al_{Int}^{0}$. When the Fermi Level equals to VBM $Al_{Ti}^{3+}$ is more stable than $Al_{Int}^{0}$ by 7.258 eV. The ε(4+/3+) and ε(3+/0) for $Al_{Ti}$ and $Al_{Int}$ is 0.519 eV and 2.192 eV respectively, indicating that the first is a deep donor while the latter is deep triple acceptor in r-TiO$_2$'s host lattice.

As $Al_{Ti}^{3+}$ is substantially more stable in n-type TiO$_2$ for both of its polymorphs the $\Delta^F$ per formula of a-TiO$_2$:Al$_{Ti}$ and r-TiO$_2$:Al$_{Ti}$ for $E_{Fermi}$ = VBM were compared. It was found that normalized $\Delta^F$(a-TiO$_2$:Al)/f.u. was lower than $\Delta^F$(r-TiO$_2$:Al)/f.u. by 17.68 meV. The results indicate that Al dopants act as weak inhibitors of the anatase to rutile phase transformation.

### TiO$_2$:Fe

Fe usually occurs in 2+ and 3+ oxidation states. $Fe_{Ti}^{2+}$ and $Fe_{Ti}^{3+}$ act as double and single acceptors in TiO$_2$ while $Fe_{Int}^{2+}$ and $Fe_{Int}^{3+}$ act as double and triple donors in TiO$_2$. As shown in **Figure 2**(e), in a-TiO$_2$, $Fe_{Ti}$ is more stable than $Fe_{Int}$ for entire range of $E_{Fermi}$ over the band gap. When the Fermi level is located at VBM, the $E^f$ of $Fe_{Ti}^{4+}$ is 0.020 eV and is slightly lower than the one of $Fe_{Int}^{3+}$ with $E^f$ of 0.021 eV. The stability of $Fe_{Ti}^{4+}$ at $E_{Fermi}$ = VBM indicates that $Fe_{Ti}$ is not electrically active in p-type TiO$_2$. The first transition level ε(4+/3+) for $Fe_{Ti}$ occurs when $E_{Fermi}$ = 0.880 eV while the second transition level ε (3+/2+) occurs when $E_{Fermi}$ = 1.315 eV. Both of these levels are located in the middle of the band gap and indicate that $Fe_{Ti}$ is a deep acceptor in a-TiO$_2$. In the case of n-type a-TiO$_2$ $Fe_{Ti}^{2+}$ with $E^f$ = -4.225 eV is more stable than $Fe_{Int}^{0}$ by 8.669 eV implying that $Fe_{Int}$ is unlikely to exist in considerable concentrations. In r-TiO$_2$, according to Figure 2(f), When the $E_{Fermi}$ equals to VBM, $Fe_{Int}^{3+}$ with $E^f$ = -2.381 eV is more stable than $Fe_{Ti}^{4+}$ of which $E^f$ = -0.372 eV. However after $E_{Fermi}$ = -0.669 eV the $E^f$ of $Fe_{Ti}$ becomes more stable than the one of $Fe_{Int}$. When $E_{Fermi}$ is located at CBM as the case of n-type TiO$_2$, the $Fe_{Ti}^{2+}$ with $E^f$ = -3.821 eV is more stable than the $Fe_{Int}^{0}$ by 7.539 eV. As a result, one can say that generally $Fe_{Ti}$ is more stable than $Fe_{Int}$ in r-TiO$_2$ for most of the allowable range of $E_{Fermi}$. Similar to the case of a-TiO$_2$, $Fe_{Ti}$ possesses two transition levels in band gap of r-TiO$_2$: ε(4+/3+) = 1.270 eV and ε(3+/2+) = 1.682 eV. These transition levels are slightly closer to the CBM when compared to the ones of a-TiO$_2$, indicating that $Fe_{Ti}$ constitutes a shallower acceptor in r-TiO$_2$.

However these transition levels are still considered so deep in the band gap that $Fe_{Ti}$ is not expected to be thermally ionized in r-TiO$_2$.

Since $Fe_{Ti}^{2+}$ is the most stable in both a- and r-TiO$_2$ when $E_{Fermi}$ = CBM as in the case of *n*-type TiO$_2$, the total energy per unit formula of these two systems were compared together. It was found that $\Delta^F$(a-TiO$_2$:Fe)/f.u. was lower than $\Delta^F$(r-TiO$_2$:Fe)/f.u. by 18.47 meV, indicating Fe dopants act as phase transformation inhibitors in TiO$_2$.

***TiO$_2$:F***

F occurs in 1- oxidation state. When F$^-$ substitutes an O$^{2-}$, it forms a single donor. While $F_{Int}^-$ constitutes a single acceptor when located in TiO$_2$ host lattice. The $E^f$ of both $F_O$ and $F_{Int}$ in a-TiO$_2$ is demonstrated in Figure 2(g). When the Fermi level is located at VBM, $F_O^-$ with $E^f$ = -1.980 eV is more stable than $F_{Int}^-$ that has $E^f$ = -0.269 eV. For $E_{Fermi}$ > 0.856 eV $F_{Int}^-$ becomes more stable than $F_O$. When $E_{Fermi}$ is located at CBM, $E^f$ of $F_{Int}^-$ and $F_O^{2-}$ equals to -3.469 eV and -0.629 eV respectively, indicating the stabilization of $F_{int}$ over $F_O$ by -2.840 eV in *n*-type a-TiO$_2$. The behaviour of F in r-TiO$_2$ is demonstrated in Figure 2(h). For $E_{Fermi}$ = VBM, $F_O^-$ with $E^f$ = -2.716 eV is substantially more stable than $F_{Int}^0$ with $E^f$ = 1.158 eV. $F_O^-$ remains the most stable form of F in r-TiO$_2$'s host lattice for most of the band gap region when its transition level ε(1-/2-) occurs at $E_{Fermi}$ = 1.825 eV where $F_O^{2-}$ becomes more stable. 2- oxidation state for $F_O^{2-}$ implies that the extra electron from neighbouring Ti ions remains localized on an F site and does not participate in conduction mechanism. For $E_{Fermi}$ > 2.621 eV, the ionized interstitial F ($F_{Int}^-$) becomes the most stable form of F in r-TiO$_2$ with its $E^f$ = -1.470 eV for $E_{Fermi}$ = CBM.

Since $F_{Int}^-$ is the most stable in both a- and r-TiO$_2$ when $E_{Fermi}$ = CBM as in the case of *n*-type TiO$_2$, the total energy per unit formula of these two systems were compared together. It was found that $\Delta^F$(a-TiO$_2$:F)/f.u. was lower than $\Delta^F$ (r-TiO$_2$:F)/f.u. by 84.78 meV, indicating F dopants act as inhibitors in TiO$_2$.

**Table 2. Stable dopant configurations and formation energies in anatase and rutile TiO$_2$ and total free energy differentials**

| System (*n*-type) | Most stable configuration in a-TiO$_2$ | Dopant's $E^f$ in a-TiO$_2$ (eV) | Most stable configuration in r-TiO$_2$ | Dopant's $E^f$ in r-TiO$_2$ (eV) | $\Delta E$ (meV) |
|---|---|---|---|---|---|
| TiO$_2$:Si | $Si_{Ti}^{2+}$ | -3.126 | $Si_{Ti}^{2+}$ | -2.386 | 32.43 |
| TiO$_2$:Al | $Al_{Ti}^{3+}$ | -2.658 | $Al_{Ti}^{3+}$ | -2.272 | 17.68 |
| TiO$_2$:Fe | $Fe_{Ti}^{2+}$ | -4.225 | $Fe_{Ti}^{2+}$ | -3.820 | 18.47 |
| TiO$_2$:F  | $F_I^-$ | -3.468 | $F_I^-$ | -1.470 | 84.78 |

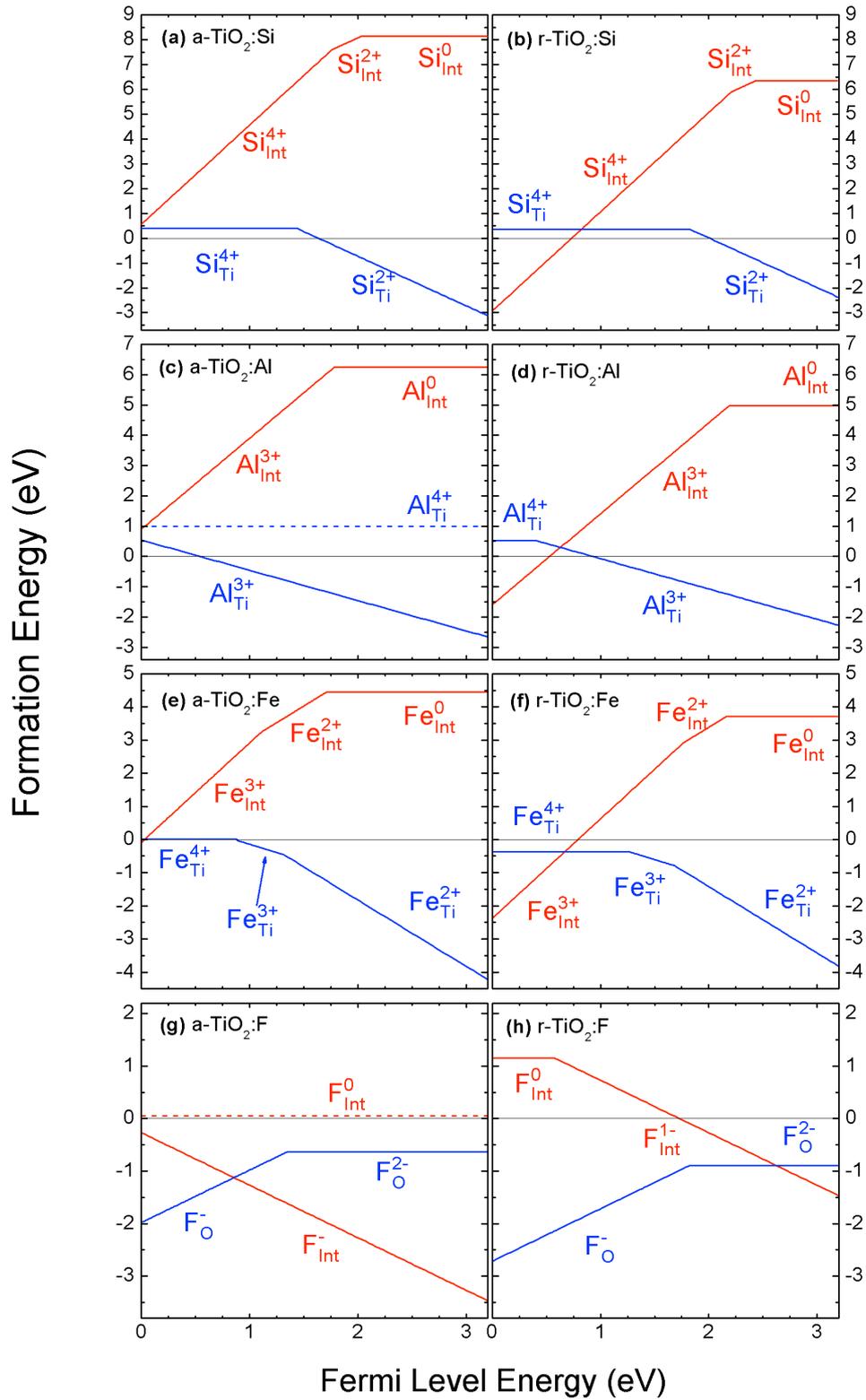

**Figure 2.** Formation energies ($E^f$) of interstitial (red) and substitutional (blue) dopants in anatase (a, c, e, g) and rutile (b, d, f, h) titanium dioxide.

## 4. Discussion

### 4.1. Pristine TiO$_2$

Previous studies have examined the phase stability of pristine anatase and rutile from experimental and first principles approaches [78]. In the present work, comparable results were obtained for the lattice parameters for these two important phases of TiO$_2$. However, thermodynamic data at room temperature differs from what has been reported to date.

A comparison of the calculated formation enthalpies of rutile and anatase yields a $\Delta E$ value of just 0.289 KJ mol$^{-1}$ for the theoretical transformation of pristine anatase to rutile at room temperature, although, as the result of an energetic barrier, this transformation does not take place at room temperature. This value suggests a narrow room temperature margin of stability and is lower than reported values to date, which have reported values between 1.7 and 12 KJ mol$^{-1}$ [74, 78-80]. The discrepancies between calculated margins of stability, in this paper and in previous reports, is not unusual and is likely the result of approximated local density functional used [78].

### 4.2. Doped TiO$_2$

The current work presents a novel Ab-initio study of dopant formation enthalpies in interstitial and substitutional positions and a total free-energy analysis of doped a-TiO$_2$ and r-TiO$_2$. From the calculations shown here it is apparent that doping with Si, Al, Fe and F should bring about a stabilization of the anatase phase.

As shown in Figure 2, from consideration of formation energies, substitutional Si$^{2+}$ doping is energetically favoured in both a-TiO$_2$ and r-TiO$_2$. This is in contrast with some reports which suggest Si doping inhibits the phase transformation by occupying interstices in TiO$_2$,[81, 82] and other reports which have suggested silicon doping occurs by substitutional Si$^{4+}$ accompanied by a lattice parameter contraction. [83] In general, the calculated inhibition of the anatase to rutile phase transformation by Si dopant, shown in Table 2, is in good agreement with reported experimental findings in the literature [62, 83-86].

While doped anatase has, in the past, been assumed to be kinetically stabilized, calculated results from the present work show that doped anatase phase TiO$_2$ is thermodynamically the most stable phase, at standard conditions of temperature and pressure, exhibiting lower $\Delta^F$.

As with Si doping, the calculated stabilization of the anatase phase by Al-doping is in agreement with previous experimental findings. The results presented here show that substitutional Al$^{3+}$ doping, which is energetically favoured over interstitial Al$^{3+}$, doping does not bring about an increase in anion vacancies rather charge compensation for Al$^{3+}$ substituting for Ti$^{4+}$ occurs through the presence of mobile charge carriers or other unintentional shallow donors. This contrasts with the previous assumptions that Al doping stabilized anatase by occupying interstices and grain boundaries, and that substitution of Ti$^{4+}$ by cations of lower valence would necessarily increase oxygen vacancy levels and enhance the phase transformation. [29, 57] This finding may help explain the observation of anatase stabilization by low valence dopants [87].

Fe doping of $TiO_2$ has been reported to bring about mixed effects on the anatase to rutile phase transformation with some reports suggesting an inhibiting effect [88] and others suggesting a promotion of the transformation [9, 56, 89]. Calculations from an energetic perspective suggest Fe is present in $TiO_2$ as substitutional $Fe^{2+}$ in Ti lattice positions and brings about a stabilization of the anatase phase. It should be noted that the computational approach considers the effect of single impurity dopant while in experimental situation inhibiting Fe ions may compete against other promoting dopants which might be the cause of observed controversy. [90-93]

The only anion dopant for which calculations were carried out in this study, was fluorine. The stabilization of anatase was strongest in this case with the order of stabilization being F>Si>Fe>Al as indicated by a comparison of $\Delta^F$(a-$TiO_2$:F) with $\Delta^F$ (r-$TiO_2$:F) which gave $\Delta E$ values of 84.78, 32.43, 18.46 and 17.68 meV respectively. The inhibition of rutile formation by F doping is supported by previously reported experimental results [94, 95].

While from energetic considerations, the studied systems are most stable with substitutional doping in the anatase phase, restricted elemental diffusion, morphological aspects and reactions of dopant compound are likely to bring about significant deviation from calculated behaviour.

## 5. Conclusions

Based on DFT calculations, all investigated cationic dopants are more stable in Ti substitutional lattice sites and are predicted to stabilise the anatase phase relative to rutile under standard conditions of pressure and temperature when present as sole-dopants.

Anionic doping with Fluorine exhibits lower formation energy in interstitial lattice sites and is predicted to bring about a significant inhibition of the anatase to rutile transformations.

The order in which anatase is stabilised was F>Si>Fe>Al in order of stabilisation strength.

Substitutional doping with cations of valence lower than 4+ does necessarily not bring about an increase in anion vacancies, rather charge compensation is predicted to occur through the formation of free holes, deep or shallow.

**Acknowledgements**
This work was supported by Australian Research Council, Grant Nos. DP1096769, DP0770424, FF0883231 and CE0348243. Computational facilities were provided by the Australian National Computational Infrastructure and INTERSECT through project db1.

## References

1.	Fujihara K, Ohno T and Matsumura M (1998) Splitting of water by electrochemical combination of two photocatalytic reactions on TiO2 particles. J Chem Soc Faraday Trans 94: 3705 - 3709
2.	Fujishima A and Honda K (1972) Electrochemical photolysis of water at a semiconductor electrode. Nature 238: 37-38
3.	Ni M, Leung M, Leung D and Sumathy K (2007) A review and recent developments in photocatalytic water-splitting using TiO2 for hydrogen production. Renewable and Sustainable Energy Reviews 11: 401-425
4.	Fujishima A, Rao TN and Tryk DA (2000) Titanium dioxide photocatalysis. Journal of Photochemistry & Photobiology, C: Photochemistry Reviews 1: 1-21
5.	Balasubramanian G, Dionysiou DD, Suidan MT, Baudin I and Lan JM (2004) Evaluating the activities of immobilized $TiO_2$ powder films for the photocatalytic degradation of organic contaminants in water. Applied Catalysis B, Environmental 47: 73-84
6.	Byrne JA, Eggins BR, Brown NMD, McKinney B and Rouse M (1998) Immobilisation of $TiO_2$ powder for the treatment of polluted water. Applied Catalysis B, Environmental 17: 25-36
7.	Hur J and Koh Y (2002) Bactericidal activity and water purification of immobilized TiO 2 photocatalyst in bean sprout cultivation. Biotechnology Letters 24: 23-25
8.	Mills A, Davies RH and Worsley D (1993) Water purification by semiconductor photocatalysis. Chemical Society Reviews 22: 417-434
9.	Carneiro JO, Teixeira V, Portinha A, Magalhaes A, Countinho P and Tavares CJ (2007) Iron-doped photocatalytic TiO2 sputtered coatings on plastics for self-cleaning applications. Materials Science and Engineering B 138: 144-150
10.	Mills A, Hodgen S and Lee SK (2004) Self-cleaning titania films: an overview of direct, lateral and remote photo-oxidation processes. Res Chem Intermed 31: 295-308
11.	Parkin IP and Palgrave RG (2005) Self-cleaning coatings. Journal of materials chemistry 15: 1689-1695
12.	Paz Y and Heller A (1997) Photo-oxidatively self-cleaning transparent titanium dioxide films on soda lime glass: The deleterious effect of sodium contamination and its prevention. J Mater Res 12: 2759
13.	Ditta I, Steele A, Liptrot C, Tobin J, Tyler H, Yates H, Sheel D and Foster H (2008) Photocatalytic antimicrobial activity of thin surface films of TiO 2, CuO and TiO 2/CuO dual layers on Escherichia coli and bacteriophage T4. Applied microbiology and biotechnology 79: 127-133
14.	Hajkova P, Spatenka P, Horsky J, Horska I and Kolouch A (2007) Photocatalytic effect of $TiO_2$ films on viruses and bacteria. Plasma Processes and Polymers 4: S397-S401
15.	Mitoraj D, Janczyk A, Strus M, Kisch H, Stochel G, Heczko PB and Macyk W (2007) Visible light inactivation of bacteria and fungi by modified titanium dioxide. Photochemical & Photobiological Sciences 6: 642-648
16.	Barzykin AV and Tachiya M (2002) Mechanism of charge recombination in dye-sensitized nanocrystalline semiconductors: Random flight model. J Phys Chem B 106: 4356-4363
17.	Gratzel M (2005) Solar energy conversion by dye-sensitized photovoltaic cells. Inorg Chem 44: 6841-6851

18. Huang SY, Schlichthorl G, Nozik AJ, Gratzel M and Frank AJ (1997) Charge recombination in dye-sensitized nanocrystalline TiO2 solar cells. J Phys Chem B 101: 2576-2582
19. O'Regan B and Gratzel M (1991) A low-cost, high-efficiency solar cell based on dye-sensitized colloidal TiO2 films.
20. Anpo M and Che M (1999) Applications of photoluminescence techniques to the characterization of solid surfaces in relation to adsorption, catalysis, and photocatalysis. Advances in catalysis 44: 119-257
21. Miyashita K, Kuroda S, Tajima S, Takehira K, Tobita S and Kubota H (2003) Photoluminescence study of electron–hole recombination dynamics in the vacuum-deposited SiO2/TiO2 multilayer film with photo-catalytic activity. Chemical Physics Letters 369: 225-231
22. Anpo M (1997) Photocatalysis on titanium oxide catalysts: approaches in achieving highly efficient reactions and realizing the use of visible light. Catalysis surveys from Japan 1: 169-179
23. Lee MC and Choi W (2002) Solid phase photocatalytic reaction on the soot/TiO2 interface: the role of migrating OH radicals. The Journal of Physical Chemistry B 106: 11818-11822
24. Lee SK, McIntyre S and Mills A (2004) Visible illustration of the direct, lateral and remote photo catalytic destruction of soot by titania. Journal of photochemistry and photobiology A 162: 203-106
25. Daude N, Gout C and Jouanin C (1977) Electronic band structure of titanium dioxide. Physical Review B 15: 3229-3235
26. Madhusudan Reddy K, Manorama SV and Ramachandra Reddy A (2003) Bandgap studies on anatase titanium dioxide nanoparticles. Materials Chemistry & Physics 78: 239-245
27. Morikawa T, Asahi R, Ohwaki T, Aoki K and Taga Y (2001) Band-gap narrowing of titanium dioxide by nitrogen doping. JAPANESE JOURNAL OF APPLIED PHYSICS PART 2 LETTERS 40: 561-563
28. Sclafani A and Herrmann JM (1996) Comparison of the photoelectronic and photocatalytic activities of various anatase and rutile forms of titania. Journal of Physical Chemistry 100: 13655 - 13661
29. Hanaor D and Sorrell CC (2011) Review of the anatase to rutile phase transformation. Journal of Materials Science 46: 855-874
30. Gopal M, Moberly Chan WJ and De Jonghe LC (1997) Room temperature synthesis of crystalline metal oxides. Journal of Materials Science 32: 6001-6008
31. Bakardjieva S, Subrt J, Stengl V, Dianez MJ and Sayagues MJ (2005) Photoactivity of anatase-rutile TiO2 nanocrystalline mixtures obtained by heat treatment of homogeneously precipitated anatase. Applied Catalysis B: Environmental 58: 193-202
32. Hanaor D, Triani G and Sorrell CC (2011) Morphology and photocatalytic activity of highly oriented mixed phase titanium dioxide thin films. Surface and Coatings Technology 205: 3658-3664
33. Hurum DC, Agrios AG, Gray KA, Rajh T and Thurnauer MC (2003) Explaining the enhanced photocatalytic activity of Degussa P25 mixed-phase TiO$_2$ using EPR. J Phys Chem B 107: 4545-4549
34. Ohno T, Sarukawa K and Matsumura M (2002) Crystal faces of rutile and anatase TiO2 particles and their roles in photocatalytic reactions. New journal of chemistry 26: 1167-1170

35. Ohno T, Tokieda K, Higashida S and Matsumura M (2003) Synergism between rutile and anatase TiO2 particles in photocatalytic oxidation of naphtalene. Applied Catalysis A 244: 383-391
36. Testino A, Bellobono IR, Buscaglia V, Canevali C, D'Arienzo M, Polizzi S, Scotti R and Morazzoni F (2007) Optimizing the photocatalytic properties of hydrothermal TiO2 by the control of phase composition and particle morphology. A systematic approach. Journal of the American Chemical Society 129: 3564-3575
37. Kominami H, Ishii Y, Kohno M, Konishi S, Kera Y and Ohtani B (2003) Nanocrystalline brookite-type titanium (IV) oxide photocatalysts prepared by a solvothermal method: correlation between their physical properties and photocatalytic activities. Catalysis Letters 91: 41-47
38. Ohtani B, Handa J, Nishimoto S and Kagiya T (1985) Highly active semiconductor photocatalyst: Extra-fine crystallite of brookite TiO2 for redox reaction in aqueous propan-2-ol and/or silver sulfate solution. Chemical physics letters 120: 292-294
39. Hu Y, Tsai HL and Huang CL (2003) Effect of brookite phase on the anatase-rutile transition in titania nanoparticles. Journal of the European Ceramic Society 23: 691-696
40. Wang H and Lewis JP (2006) Second-generation photocatalytic materials: anion-doped $TiO_2$. Journal of Physics, Condensed Matter 18: 421-434
41. Hanaor D, Michelazzi M, Chenu J, Leonelli C and Sorrell CC (2011) The effects of firing conditions on the properties of electrophoretically deposited titanium dioxide films on graphite substrates. Journal of the European Ceramic Society 31: 2877-2885
42. Liu G, Wang L, Yang HG, Cheng HM and Lu GQM (2009) Titania-based photocatalysts—crystal growth, doping and heterostructuring. J Mater Chem 20: 831-843
43. Kudo A and Miseki Y (2008) Heterogeneous Photocatalyst Materials For Water Splitting. Chemical Society Reviews 38: 253-278
44. Burns A, Li W, Baker C and Shah SI (2002) Sol-Gel Synthesis and Characterization of Neodymium-Ion Doped Nanostructured Titania Thin Films. Materials Research Society Symposium Proceedings 703: 5.2.1-5.2.6
45. Batzill M, Morales EH and Diebold U (2006) Influence of Nitrogen Doping on the Defect Formation and Surface Properties of TiO2 Rutile and Anatase. Physical review letters 96: 26103
46. Xin B, Ren Z, Wang P, Jing L, Fu H and Liu J (2007) Study on the mechanisms of photoinduced carriers separation
and recombination for $Fe^{3+}$ TiO2 photocatalysts. Applied Surface Science 253: 4390-4395
47. Sun B, Vorontsov AV and Smirniotis PG (2003) Role of Platinum Deposited on $TiO_2$ in Phenol Photocatalytic Oxidation. Langmuir 19: 3151-3156
48. Nagaveni K, Hegde MS, Ravishankar N, Subbanna GN and Madrass G (2004) Synthesis and structure of nanocrystalline TiO2 with lower band gap showing high photocatalytic activity. Langmuir 20: 2900-2907
49. Serpone N (2006) Is the Band Gap of Pristine $TiO_2$ Narrowed by Anion-and Cation-Doping of Titanium Dioxide in Second-Generation Photocatalysts? J Phys Chem B 110: 24287-24293
50. Umebayashi T, Yamaki T, Itoh H and Asai K (2002) Band gap narrowing of titanium dioxide by sulfur doping. Applied Physics Letters 81: 454
51. Baiju KV, Sibu CP, Rajesh K, Pillai PK, Mukundan P, Warrier KGK and Wunderlich W (2005) An aqueous sol–gel route to synthesize nanosized lanthana-doped titania having an increased anatase phase stability for photocatalytic application. Materials Chemistry & Physics 90: 123-127

52. Kim D, Kim T and Hong K (1999) Low-firing of CuO-doped anatase. Materials Research Bulletin 34: 771-781
53. Kim J, Song KC, Foncillas S and Pratsinis S (2001) Dopants for synthesis of stable bimodally porous titanis. Journal of the European Ceramic Society 21: 2863 - 2872
54. Reidy DJ, Holmes JD, Nagle C and Morris MA (2005) A highly thermally stable anatase phase prepared by doping with zirconia and silica coupled to a mesoporous type synthesis technique. Journal of Materials Chemistry: 3494-3500
55. Sharma SD, Singh D, Saini K, Kant C, Sharma V, Jain SC and Sharma CP (2006) Sol-gel derived super-hydriphilic nickel doped $TiO_2$ film as an active photocatalyst. Applied Catalysis A 314
56. Mackenzie KJD (1975) Calcination of titania V. Kinetics and mechanism of the anatase-rutile transformation in the presence of additives. Transactions and Journal of the British Ceramic Society 74: 77-84
57. Shannon RD and Pask JA (1965) Kinetics of the anatase-rutile transformation. Journal of the American Ceramic Society 48: 391-398

58. Kohn W, Sham LJ and JOLLA. CUSDL (1965) Self-consistent equations including exchange and correlation effects APS
59. Sousa SF, Fernandes PA and Ramos MJ (2007) General performance of density functionals. Journal of Physical Chemistry A 111: 10439-10452
60. Parr RG and Yang W (1994) Density-functional theory of atoms and molecules Oxford University Press, USA
61. Delley B (1990) An all-Electron Numerical Method for Solving the Local Density Functional for Polyatomic Molecules. Journal of Chemical Physics 92: 508-517
62. Delley B (2000) From molecules to solids with the DMol(3) approach. Journal of Chemical Physics 113: 7756-7764
63. Perdew JP and Wang Y (1992) Accurate and Simple Analytic Representation of the Electron Gas Correlation Energy. Physical Review B 45: 13244-13249
64. Becke AD (1993) Density-Functional Thermochemistry III. The Role of Exact Exchange. The Journal of Chemical Physics 98: 5648-5652
65. Bloch F (1929) Über die quantenmechanik der elektronen in kristallgittern. Zeitschrift für Physik A Hadrons and Nuclei 52: 555-600
66. Monkhorst HJ and Pack JD (1976) Special points for Brillouin-zone integrations. Physical Review B 13: 5188-5192
67. Zhu Z, Shima N and Tsukada M (1989) Electronic states of Si (100) reconstructed surfaces. Physical Review B 40: 11868
68. Shanno D and Kettler P (1970) Optimal conditioning of quasi-Newton methods. Math Comp 24: 657-664
69. Hine NDM, Robinson M, Haynes PD, Skylaris CK, Payne MC and Mostofi AA (2011) Accurate ionic forces and geometry optimization in linear-scaling density-functional theory with local orbitals. Physical Review B 83: 195102
70. Godbout N, Salahub DR, Andzelm J and Wimmer E (1992) Optimization of Gaussian-type basis sets for local spin density functional calculations. Part I. Boron through neon, optimization technique and validation. Canadian Journal of Chemistry 70: 560-571
71. Curtiss LA, Carpenter JE, Raghavachari K and Pople JA (1992) Validity of Additivity Approxiamations in Gaussian Theory. Journal of Chemical Physics 96: 9030-9034
72. Cromer DT and Herrington K (1955) The Structures of Anatase and Rutile. Journal of the American Chemical Society 77: 4708-4709

73. Na-Phattalung S, Smith MF, Kim K, Du MH, Wei SH, Zhang SB and Limpijumnong S (2006) First-principles study of native defects in anatase $TiO_2$. Physical Review B 73
74. Smith SJ, Stevens R, Liu S, Li G, Navrotsky A, Boerio-Goates J and Woodfield BF (2009) Heat capacities and thermodynamic functions of $TiO_2$ anatase and rutile: Analysis of phase stability. American Mineralogist 94: 236
75. Zhang H and Banfield JF (1998) Thermodynamic analysis of phase stability of nanocrystalline titania. Journal of Materials Chemistry 8: 2073-2076
76. Van de Walle CG and Neugebauer J (2004) First-principles calculations for defects and impurities: Applications to III-nitrides. Journal of applied physics 95: 3851-3879
77. Janotti A and Van de Walle CG (2006) Hydrogen multicentre bonds. Nature materials 6: 44-47
78. Muscat J, Swamy V and Harrison NM (2002) First-principles calculations of the phase stability of TiO2. Physical Review B 65: 224112-16
79. Stull DR and Prophet H (1971) JANAF thermochemical tables. NBS STP NO 37, WASHINGTON DC, 1971, 1141 P
80. Navrotsky A and Kleppa OJ (1967) Enthalpy of the Anatase-Rutile Transformation. Journal of the American Ceramic Society 50: 626-626
81. Akhtar MK, Pratsinis SE and Mastrangelo SVR (1992) Dopants in Vapor-Phase Synthesis of Titania Powders. Journal of the American Ceramic Society 75: 3408-3416
82. Chen CH, Kelder EM and Schoonman J (1999) Electrostatic sol-spray deposition (ESSD) and characterisation of nanostructured TiO2 thin films. Thin Solid Films 342: 35-41
83. Okada K, Yamamoto N, Kameshima Y and Yasumori A (2001) Effect of Silica Additive on the Anatse to Rutile Phase Transition. Journal of the American Ceramic Society 84: 1591-1596
84. Reidy DJ, Holmes JD and Morris MA (2006) Preparation of a highly thermally stable titania anatase phase by addition of mixed zirconia and silica dopants. Ceramics International 32: 235-239
85. Zhang YH and Reller A (2002) Phase transformation and grain growth of doped nanosized titania. Materials Science & Engineering C 19: 323-326
86. Yang J and Ferreira JMF (1998) Inhibitory effect of the Al2O3 -SiO2 mixed additives on the anatase-rutile phase transformation. Materials letters(General ed) 36: 320-324
87. Vargas S, Arroyo R, Haro E and Rodriguez R (1999) Effect of cationic dopants on the phase transition temperature of titania prepared by the Sol-gel method. J Mater Res 14: 3932-3937
88. Janes R, Knightley LJ and Harding CJ (2004) Structural and spectroscopic studeis of iron (iii) doped titania powders. Dyes and Pigments 62: 199-212
89. Iida Y and Ozaki S (1961) Grain Growth and Phase Transformation of Titanium Oxide During Calcination. Journal of the American Ceramic Society 44: 120-127
90. MacKenzie KJD (1975) Calcination of titania IV. Effect of additives on the anatase-rutile transformation. Transactions and Journal of the British Ceramic Society 74: 29-34
91. Wang ZL, Yin JS, Mo WD and Zhangs ZJ (1997) In-situ analysis of valence conversion in transition metal oxides using electron energy-loss spectroscopy. J Phys Chem B 101: 6793-6798
92. Heald EF and Weiss CW (1972) Kinetics and mechanism of the anatase/rutile transformation as catalyzed by ferric oxide and reducing conditions. American Mineralogist 57: 10-23
93. Gennari FC and Pasquevich DM (1998) Kinetics of the anatase rutile transformation in TiO2 in the presence of Fe2O3. Journal of Materials Science 33: 1571-1578

94.     Takahashi Y and Matsuoka Y (1988) Dip-coating of $TiO_2$ films using a sol derived from Ti (O-i-Pr) 4-diethanolamine-$H_2$O-i-PrOH system. Journal of Materials Science 23: 2259-2266

95.     Yu JC, Yu J, Ho W, Jiang Z and Zhang L (2002) Effects of F-doping on the photocatalytic activity and microstructures of nanocrystalline $TiO_2$ powders. Chem Mater 14: 3808-3816